# The Rise of Blockchain Technology in Agriculture and Food Supply Chains

**Andreas Kamilaris[1,2], Agusti Fonts[1] and Francesc X. Prenafeta-Boldύ[1]**

[1] GIRO Program, IRTA Torre Marimon, E-08140 Caldes de Montbui, Barcelona, Spain

[2] Research Centre on Interactive Media, Smart Systems and Emerging Technologies (RISE), Nicosia, Cyprus

**Abstract:**

Blockchain is an emerging digital technology allowing ubiquitous financial transactions among distributed untrusted parties, without the need of intermediaries such as banks. This article examines the impact of blockchain technology in agriculture and food supply chain, presents existing ongoing projects and initiatives, and discusses overall implications, challenges and potential, with a critical view over the maturity of these projects. Our findings indicate that blockchain is a promising technology towards a transparent supply chain of food, with many ongoing initiatives in various food products and food-related issues, but many barriers and challenges still exist, which hinder its wider popularity among farmers and systems. These challenges involve technical aspects, education, policies and regulatory frameworks.

**Keywords:** Blockchain Technology, Digital Agriculture, Food Supply Chain, Barriers, Benefits, Challenges.

## 1. Introduction

A decade has passed since the release of the whitepaper "Bitcoin: A Peer-to-Peer Electronic Cash System" by the pseudonymous author (Nakamoto 2008). This work set basis for the development of Bitcoin, the first cryptocurrency that allowed reliable financial transactions without the need of a trusted central authority, such as banks and financial institutions (Tschorsch and Scheuermann 2016). Bitcoin solved the double-spending problem (i.e. the flaw associated to digital tokens because, as computer files, can easily be duplicated or falsified), with the invention of the blockchain technology. A blockchain is a digital transaction ledger, maintained by a network of multiple computing machines that are not relying on a trusted third party. Individual transaction data files (blocks) are managed through specific software platforms that allow the data to be transmitted, processed, stored, and represented in human readable form. In its original bitcoin configuration, each block contains a header with a time-stamp, transaction data and a link to the previous block. A hash gets generated for every block, based on its contents, and then becomes referred in the heading of the subsequent block (see Figure 1). Hence, any manipulation of a given block would result in a mismatch in the hashes of all successive blocks.

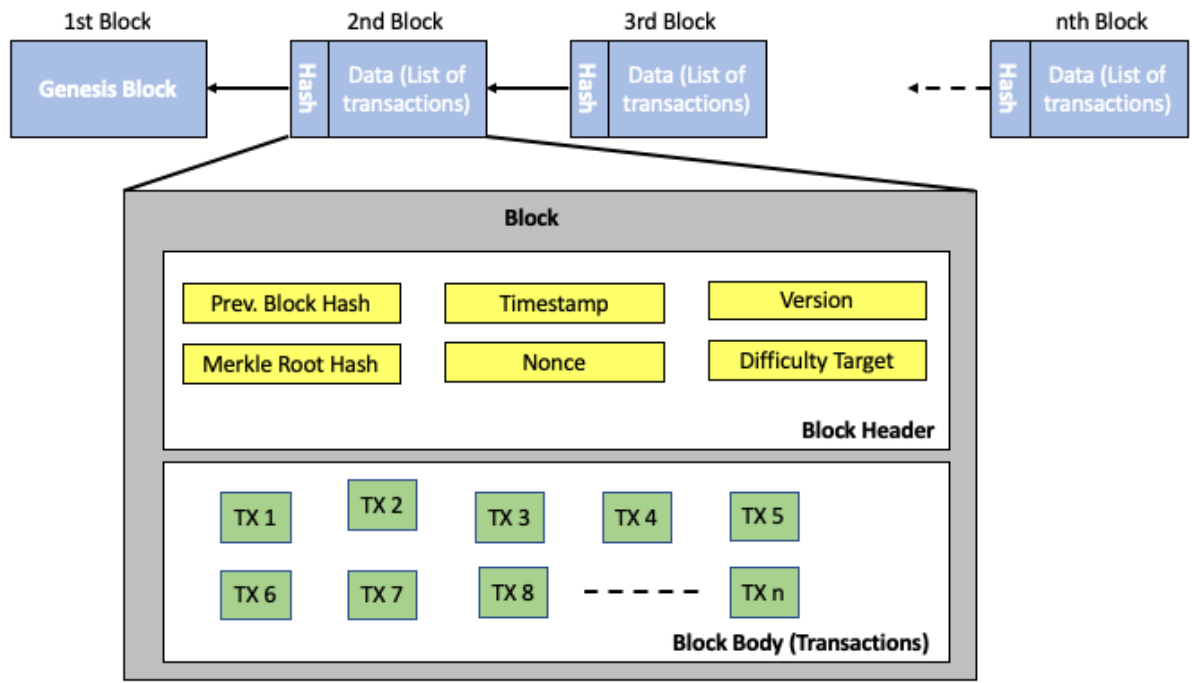

Figure 1: Example of a blockchain containing *n* blocks, in which each successive block contains the hash of the previous block, a timestamp, the transaction information, the nonce number for the mining process and other details needed for the protocol to work.

Every transaction is disseminated through the network of machines running the blockchain protocol, and needs to be validated by all computer nodes. The key feature of a blockchain is its ability to keep a consistent view and agreement among the participants (i.e. *consensus*) (Bano 2017), even if some of them might not be honest (Castro and Liskov 1999). The problem of consensus has been extensively studied by researchers in the past, however its use in the domain of blockchain has given new stimuli and motivation, leading to novel proposals for design of blockchain systems. The most well-known, used in Bitcoin, is called "Proof of Work" (PoW) and it requires computer nodes, called miners in this case, to solve difficult computational tasks before validating transactions and be able to add them to the blockchain (Bentov, Gabizon and Mizrahi 2016). The first miner to solve the puzzle bundles the block to the chain, which is then validated by the rest, and gets rewarded with newly minted coins plus a small transaction fee.

Common criticism of the PoW include that miners compete continuously in computer power, which leads to increased hardware and energy costs, with the subsequent risks of centralization and high environmental footprint (Becker, et al. 2013), (Krause and Tolaymat 2018). An alternative consensus approach gaining momentum is called "Proof of Stake" (PoS), and it is about giving the decision-making power to entities who possess coins within the system, putting them "on stake" during transaction approval (Bentov, Gabizon and Mizrahi 2016). In PoS, the nodes are known as the 'validators' and, rather than mining the blockchain, they validate the transactions to earn a transaction fee. There is no mining to be done, as all coins exist from day one. Simply put, nodes are randomly selected to validate blocks, and the probability of this random selection depends on the amount of stake held. Consequently, PoS achieves the same effect of mining (distributed consensus) without the need of expending large amounts of computing power and energy (BitFury Group 2015). Other consensus mechanisms include Proof of Elapsed Time (PoET), Simplified Byzantine Fault Tolerance (SBFT), and Proof of Authority (PoA).

Hundreds of alternative digital tokens have appeared in the wake of this development, aiming to address some particular weaknesses of the dominant cryptocurrencies, or target a specific domain, such as health, gambling, insurance, agriculture and many others (Coinmarketcap 2017). Blockchain is also being investigated (and in some cases adopted) by the conventional banking system, and nearly 15% of financial institutions are currently using this technology for their transactions (IBM 2017).

Since 2014 it has increasingly been realized that blockchain can be used for much more than cryptocurrency and financial transactions, so that several new applications are being explored (Tayeb and Lago 2018): handling and storing administrative records, digital authentication and signature systems, verifying and tracking ownership of intellectual property rights and patent systems, enabling smart contracts, tracking patient health records, greater transparency in charities, frictionless real-estate transfers, electronic voting, distribution of locally produced goods and, in general, for tracking products as they pass through a supply chain from the manufacturer and distributor, to the final buyer. Such changes are already revolutionizing many aspects of business, government and society in general, but they might also pose new challenges and threads that need to be anticipated. Many of these new applications combine blockchain and distributed ledger technologies (DLTs) with smart contracts and decentralized applications, making third party tampering or censorship virtually impossible (Buterin 2015).

## 2. Food Supply Chain

The food chain worldwide is highly multi-actor based and distributed, with numerous different actors involved, such as farmers, shipping companies, wholesalers and retailers, distributors, and groceries. The main phases characterizing a generic agri-food supply chain are described below (Caro, Ali, et al. 2018):

1. *Production*: The production phase represents all agricultural activities implemented within the farm. The farmer uses raw and organic material (fertilizers, seeds, animal breeds and feeds) to grow crops and livestock. Throughout the year, depending on the cultivations and/or animal production cycle, we can have one or more harvest/yield.
2. *Processing*: This phase concerns the transformation, total or partial, of a primary product into one or more other secondary products. Subsequently a packaging phase is expected, where each package might be uniquely identified through a production batch code containing information such as the production day and the list of raw materials used.
3. *Distribution*: Once packaged and labeled, the product is released for the distribution phase. Depending on the product, delivery time might be set within a certain range and there might be a product storage step (Storage).
4. *Retailing*: At the end of the distribution, the products are delivered to retailers who perform the sale of the product (Retailers). The end-user of the chain will be the customer, who will purchase the product (Customer).
5. *Consumption*: The consumer is the end user of the chain, he/she buys the product and demands traceable information on quality standards, country origin, production methods, etc.

Figure 2 (top section, physical flow) illustrates a simplified version of the food supply system and its main phases and actors. This current system is till date inefficient and unreliable (Tripoli and Schmidhuber 2018). Exchange of good are based on complex and paper-heavy settlement processes while these processes are not much transparent, with high risks between buyers and sellers during exchange of value. As transactions are vulnerable to fraud, intermediaries get involved, increasing the overall costs of the transfers (Lierow, Herzog and Oest 2017). It is estimated that the cost of operating supply chains makes up two thirds of the final cost of goods. Thus, there is much space for optimization of the supply chains, by effectively reducing the operating costs. Finally, when people buy products locally, they are not aware of the origins of these goods, or the environmental footprint of production.

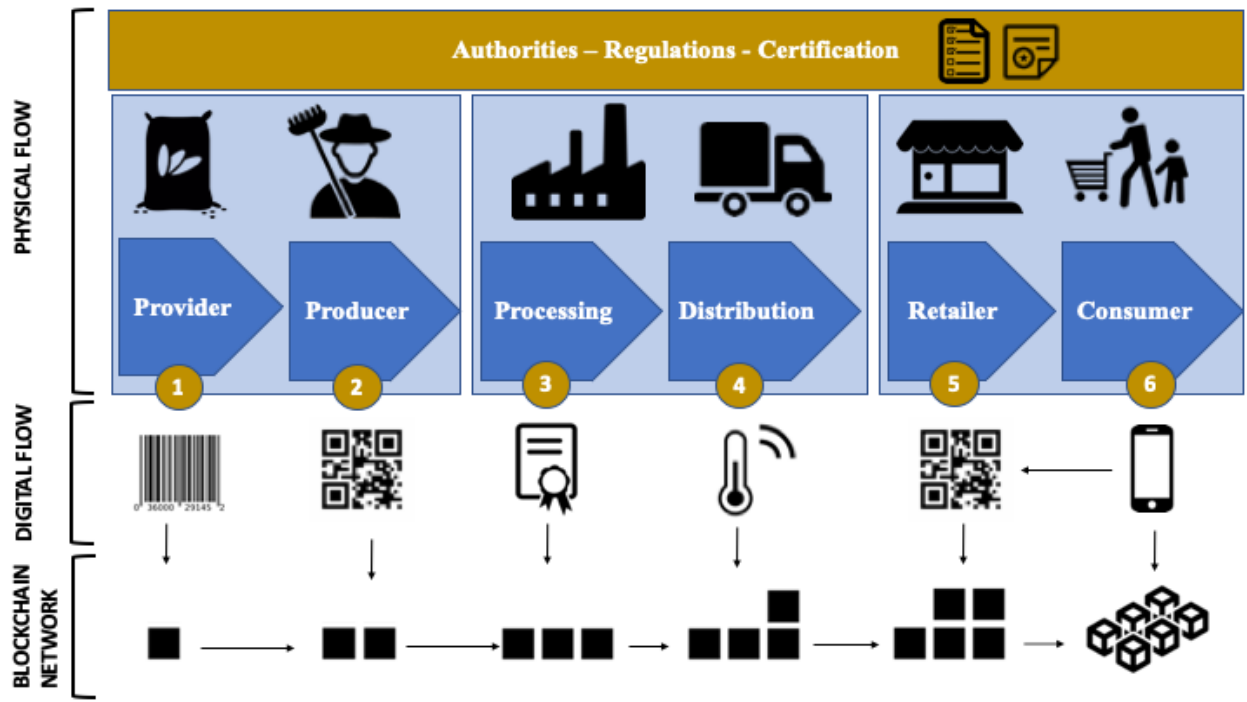

Figure 2: A simplified food supply chain system.

## 3. Blockchain in Agriculture and Food Supply Chain

While the blockchain technology gains success and proves its functionality in many cryptocurrencies, various organizations and other entities aim at harnessing its transparency and fault tolerance in order to solve problems in scenarios where numerous untrusted actors get involved in the distribution of some resource (Manski 2017), (Sharma 2017). Two important, highly relevant areas are *agriculture* and *food supply chain* (Dujak and Sajter 2019), (Tripoli and Schmidhuber 2018). Agriculture and food supply chains are well interlinked, since the products of agriculture almost always are used as inputs in some multi-actor distributed supply chain, where the consumer is usually the final client (Maslova 2017).

There is evidence that blockchain applications started to become used in the supply chain management soon after the technology appeared (Tribis, El Bouchti and Bouayad 2018). Blockchain in supply chain management is expected to grow at an annual growth rate of 87% and increase from $45 million in 2018 to $3,314.6 million by 2023 (Chang, Iakovou and Shi 2019).

As a successful example, in December 2016, the company AgriDigital executed the world's first settlement of the sale of 23.46 tons of grain on a blockchain (ICT4Ag 2017). Since then, over 1,300 users and more than 1.6 million tons of grain has been transacted over the cloud-based system, involving $360 million in grower payments. The success of AgriDigital served

as an inspiration for the potential use of this technology in the agricultural supply chain. AgriDigital is now aiming to build trusted and efficient agricultural supply chains by means of blockchain technology (AgriDigital 2017). As another recent example, Louis Dreyfus Co (LDC), one of the world's biggest foodstuffs traders, teamed up with Dutch and French banks for the first agricultural commodity trade (i.e. a cargo of soybeans from the US to China) based on blockchain (Hoffman and Munsterman 2018). According to LDC, by automatically matching data in real time, avoiding duplication and manual checks, document processing was reduced to a fifth of the time.

A simplified example of the digitization of the food supply chain, supported by blockchain technology is depicted in Figure 2. Under the physical flow (top layer), there is the digital flow layer (middle layer), consisting of various digital technologies (i.e. QR codes, RFID, NFC, online certification and digital signatures, sensors and actuators, mobile phones etc.). The Internet/Web serves as the connecting infrastructure. Every action performed along the food chain, empowered by the use of the aforementioned digital technologies, is recorded to the blockchain (bottom layer of Figure 2), which serves as the immutable means to store information that is accepted by all participating parties. The information captured during each transaction is validated by the business partners of the food supply network, forming a consensus between all participants. After each block becomes validated, it is added to the chain of transactions (as Figure 2 shows), becoming a permanent record of the entire process. At every stage of the trajectory of food (defined with numbers 1-6 in Figure 2), different technologies are involved and different information is written to the blockchain, as described below for each of these stages:

1. Provider: Information about the crops, pesticide and fertilizers used, machinery involved etc. The transactions with the producer/farmer are recorded.
2. Producer: Information about the farm and the farming practices employed. Additional info about the crop cultivation process, weather conditions, or animals and their welfare is also possible to be added.
3. Processing: Information about the factory and its equipment, the processing methods used, batch numbers etc. The financial transactions that take place with the producers and also with the distributors are recorded too.
4. Distribution: Shipping details, trajectories followed, storage conditions (e.g. temperature, humidity), time in transit at every transport method etc. All transactions between the distributors and also with the final recipients (i.e. retailers) are written on the blockchain.
5. Retailer: Detailed information about each food item, its current quality and quantity, expiration dates, storage conditions and time spent on the shelf are listed on the chain.
6. Consumer: At the final stage, the consumer can use a mobile phone connected to the Internet/Web or a web application in order to scan a QR code associated with some food item, and see in detail all information associated with the product, from the producer and provider till the retail store.

In this section of the paper, various initiatives have been identified where blockchain technology could be used to solve real-life practical problems at the agricultural supply chain. To identify relevant initiatives, a keyword-based search was performed through the web scientific indexing services *Web of Science* and *Google Scholar*. The following query was used:

*Blockchain AND [Agriculture OR Food OR "Food Supply" OR "Food Supply Chain"]*.

Our focus was on *existing* initiatives, projects and case studies, and not on the general potential of blockchain in the field. Based on this search, only 29 papers were identified. From these papers, just 23 were relevant, in terms of using blockchain technology in food supply chain. To increase bibliography, related work of the initial 29 papers was examined, together with a keyword-based search in popular search engines, increasing the number of relevant identified initiatives to 49. Based on their purpose and overall target/goal, these 49 initiatives were divided into six main categories, as follows:

a) food security (2 projects/initiatives, 4%),
b) food safety (3 projects/initiatives, 6%),
c) food integrity (24 projects/initiatives, 49.5%),
d) support of small farmers (8 projects/initiatives, 16%),
e) waste reduction and environmental awareness (5 projects/initiatives, 10%), and
f) better supervision and management of the supply chain (7 projects/initiatives, 14.5%).

An analysis of the findings is performed in Section 4. Some of the potential benefits of blockchain are listed in Section 5, while various challenges and barriers for wider adoption are identified and discussed in Section 6.

**3.1 Food Security**

The Food and Agriculture Organization (FAO) defines food security as the situation when "*all people, at all times, have physical, social and economic access to sufficient, safe and nutritious food that meets their dietary needs and food preferences for an active and healthy life*". Achieving this objective has proven to be extremely challenging under humanitarian crises related to environmental disasters, violent political and ethnic conflicts, etc. Blockchain is regarded as an opportunity for the transparent delivery of international aid, for disintermediating the process of delivery, for making records and assets verifiable and accessible and, ultimately, to respond more rapidly and efficiently in the wake of humanitarian emergencies (AID Tech 2017). Examples include digital food coupons having been distributed to Palestinian refugees in the Jordan's Azraq camp (Blockchain for Zero Hunger 2017), via an Ethereum-based blockchain (Ethereum 2015), where the coupons could be redeemed via biometric data (Built to Adapt 2018). At the moment, the project is helping 100,000 refugees.

### 3.2. Food Safety

Food safety is the condition of processing, managing and storing food in hygienic ways, in order to prevent illnesses from occurring to human population. Food safety and quality assurance have become increasingly difficult in times of growing global flows of goods (Creydt en Fischer 2019). The Center for Disease Control and Prevention (CDC) claims that contamination because of food causes 48M Americans to become ill and 3,000 to die every year (CDC 2018), (Tripoli and Schmidhuber 2018). In 2016, Oceana performed a research on seafood fraud, showing that 20% of seafood is labelled incorrectly (Oceana 2013). Lee et al. commented that food supply chains are characterized by reduced trust, long shipment distances, high complexity, and large processing times (Lee, et al. 2017). Blockchain could provide an efficient solution in the urgent need for an improved traceability of food regarding its safety and transparency. As Figure 2 shows, recording information about food products at every stage of the supply chain allows to ensure good hygienic conditions, identifying contaminated products, frauds and risks as early as possible.

Walmart and Kroger are among the first companies to embrace blockchain and include the technology into their supply chains (CB Insights 2017), working initially on case studies that focus on Chinese pork and Mexican mangoes (Kamath 2018). Early results from the studies showed that, when tracking a package of mangoes from the supermarket to the farm where they were grown, it took 6.5 days to identify the origin and the path the fruit followed with traditional methods, whereas with blockchain this information was available in just a few seconds (Wass 2017).

The integration of blockchain with Internet of Things (IoT) for real-time monitoring of physical data and tracing based on the hazard analysis and critical control points system (HACCP) has recently been proposed (Tian 2017). This is particularly critical for the maintenance of the cold-chain in the distribution logistics of spoilable food products. As an example, ZetoChain performs environmental monitoring at every link of the cold chain, based on IoT devices (Zeto 2018). Problems are identified in real-time and the parties involved are notified immediately for fast action taking. Smart contracts are harnessed to increase the safety of sales and deliveries of goods. Mobile apps can be used by consumers to scan *Zeto labels* on products in order to locate the product's history.

**3.3 Food Integrity**

Food integrity is about reliable exchange of food in the supply chain. Each actor should deliver complete details about the origin of the goods. Examples of these details have been listed at the beginning of Section 3, and the process is described in Figure 2. This issue is of great concern in China, where the extremely fast growth has created serious transparency problems (Tian 2017), (Tse, et al. 2017). Food safety and integrity can be enhanced through higher traceability (Galvez, Mejuto and Simal-Gandara 2018), (Creydt en Fischer 2019). By means of blockchain, food companies can mitigate food fraud by quickly identifying and linking outbreaks back to their specific sources (Levitt 2016). Recent research has predicted that the food traceability market will be worth $14 billion by 2019 (MarketsandMarkets Research 2016). There are numerous examples of companies, start-ups and initiatives aiming to improve food supply chain integrity through the blockchain technology. The most important on-going projects are listed below, based on their scale, their potential impact and the significance of the partners, organizations and/or actors involved.

The agricultural conglomerate Cargill Inc. aims to harness blockchain to let shoppers trace their turkeys from the store to the farm that raised them (Bunge 2017). Turkeys and animal welfare are considered at a recent pilot involving blockchain (Hendrix Genetics 2018). The European grocer Carrefour is using blockchain to verify standards and trace food origins in various categories, covering meat, fish, fruits, vegetables and dairy products (Carrefour 2018).

Downstream beer (Ireland Craft Beers 2017) is the first company in the beer sector to use blockchain technology, revealing everything one wants to know about beer, i.e. its ingredients and brewing methods. Every aspect of this craft beer is being recorded and written to the blockchain as a guarantee of transparency and authenticity. Consumers can use their smart phones to scan the QR code on the front of the bottle and they are then taken to a website where they can find relevant information, from raw ingredients to the bottling.

Concerning meat production, "Paddock to plate" is a research project aiming to track beef along the chain of production-consumption, increasing the reputation of Australia for high quality (Campbell 2017). The project uses BeefLedger as its technology platform (BeefLedger Limited 2017). As another example, the e-commerce platform JD.com monitors the beef produced in inner Mongolia, distributed to different provinces of China (JD.com Blog 2018). By scanning QR codes, one can see details about the animals involved, their nutrition, slaughtering and meat packaging dates, as well as the results of food safety tests. To guarantee to customers that its chickens are actually free-range, the Gogochicken company uses an ankle bracelet to monitor the chickens' movements and behavior via GPS tracking, and this information is then available through the web (Adele Peter, Fast Company 2017). The aim of the company is to build trust by documenting the origins of the food. Right now, 100,000 birds have been outfitted with GPS bracelets, but the Shanghai-based company plans to incorporate about 23 million birds into project over the next three years.

The Grass Roots Farmers Cooperative (Grass Roots Farmers' Cooperative 2017) sells a meat subscription box, which uses blockchain technology to inform consumers in a reliable way about the raising conditions of their animals. In the pilot performed, cases of chicken distributed in San Francisco are labeled with QR codes that link to the story of the meat they contain.

Moreover, in April 2017, Intel demonstrated how Hyperledger Sawtooth (Hyperledger 2018), a platform for creating and managing blockchains, could facilitate traceability at the seafood supply chain. The study used sensory equipment to record information about fish location and storing conditions. Hyperledger is one of the most important initiatives, based on completeness and quality of services and tools, as well as the size of the supporting community and the significance of the members that support the overall project. Hyperledger aims to offer complete solutions towards the business use of the blockchain, and it has been proposed in recent research efforts such as AgriBlockIoT (Caro, Ali, et al. 2018). Hyperledger focuses to the creation of open source frameworks based on the DLT, suitable for enterprise solutions. Two of the most mature Hyperledger frameworks are named *Fabric* (for permissioned blockchain networks) and *Sawtooth* (for both permissioned and permissionless blockchain networks). These two frameworks constitute generic enterprise-grade software, offering support for various smart contract languages and they are used by a wide community of companies, developers and users. In particular, Hyperledger Fabric is backed by IBM. While Hyperledger Fabric is the most well-known and widespread, Sawtooth is the most advanced and heavy-duty, allowing adequate integration with other blockchain frameworks (Suprunov 2018).

In January 2018, the World Wildlife Foundation (WWF) announced the Blockchain Supply Chain Traceability Project (WWF 2018), to eliminate illegal tuna fishing by means of blockchain. Through the project, fishermen can register their catch on the blockchain through RFID e-tagging and scanning fish. Traceability of tuna is also the focus of Balfegó (Balfegó Group 2017).

Furthermore, ripe.io has created the Blockchain of Food (Ripe.io 2017), which constitutes a food quality network that maps the food's journey from production to our plate. Ripe.io has recently raised $2.4 million in seed funding in a round led by the venture arm of global container logistics company Maersk (AgFunder News 2018).

Via the services provided by the OriginTrail company, consumers can see from which orchard the ingredients they cook have grown, the origin and growing conditions of poultry etc. (OriginTrail 2018). Also, the project "blockchain for agri-food" developed a proof-of-concept blockchain-based application about table grapes from South Africa (Ge, et al. 2017). A framework for greenhouse farming with enhanced security, based on blockchain technology, is proposed in (Patil, et al. 2017). Nestle has recently entered the IBM Food Trust partnership towards food traceability (ITUNews 2018), with a pilot based on canned pumpkin and mango.

Some research initiatives proposed the combination of blockchain with other technologies (i.e. IoT, RFID, NFC), in order to increase food traceability. A system based on combining RFID and blockchain technologies is discussed in (Tian 2016) while a system based on IoT devices and smart contracts is proposed in (Kim, et al. 2018). Boehm et al. proposed an updated traceability system using blockchain technology combined with Near Field Communication (NFC) and verified users (Boehm, Kim and Hong 2017).

Finally, the blockchain technology is also being assessed to trace the production of non-edible crops that are also very sensitive to integrity issues because of regulation and legal aspects. Figorilli et al. (2018) experiment with an implementation of blockchain for the electronic traceability of wood from standing tree to final user, based on RFID sensors and open source technology (Figorilli, et al. 2018). Canada is currently developing a permissioned blockchain network for the tracking of the cannabis supply chain (Abelseth 2018). By tracking the cannabis chain, Health Canada aims to enforce regulations more easily.

### 3.4 Small Farmers Support

Small cooperatives of farmers is a way to raise competitiveness in developing countries (Chinaka 2016). Via cooperatives, individual farmers are able to win a bigger share of the value of the crops they are cultivating (FarmShare 2017). FarmShare aims to create new forms of ownership of property, cooperation of communities and self-sufficient local economies. It constitutes an evolution of the community-supported agriculture model, taking advantage of the blockchain's potential for distributed consensus, token-based equity shares and automated governance in order to foster greater community engagement while removing some of the managerial burdens (FarmShare 2017).

AgriLedger uses distributed crypto-ledger to increase trust among small cooperatives in Africa (AgriLedger 2017). The authors in (Davcev, et al. 2018) proposed a new approach that leads to trusted cooperative applications and services within the agro-food chain, among farmers and other entities of the chain. OlivaCoin is a B2B platform for trade of olive oil, supporting the olive oil market, in order to reduce overall financial costs, increase transparency and gain easier access to global markets (OlivaCoin 2016).

Further, some startups support small farmers by offering tools that increase the traceability of goods, such as Provenance, Arc-Net, Bart.Digital and Bext360. As a recent example, the Soil Association Certification (Soil Association Certification 2018) has teamed up with Provenance to pilot technology which tracks the journey of organic food.

We note here that even medium-size farmers could benefit from blockchain and the aforementioned initiatives, as they form a clearly different category than the large corporations (FarmShare 2017). Cooperatives, on the other hand, might be formed by either small- or medium-size farmers, and can become quite large entities representing tens or hundreds of farmers. Blockchain could be very useful for such cooperatives, because the transparency of information involved could help to solve disputes and conflicts among the farmers in a fairer way for everyone (Chinaka 2016), (AgriDigital 2017). An example of how blockchain technology could be used for an automatic transaction between a cooperative of farmers (i.e. producers) and a distributor/retailer, via the use of smart contracts, is provided in Figure 3. The figure presents a hypothetical scenario in which a cooperative based in Africa uses a smart contract to facilitate the sale of its cereals' production. The execution of the contract involves the automatic access of the buyer to a storage room, where the crops are stored.

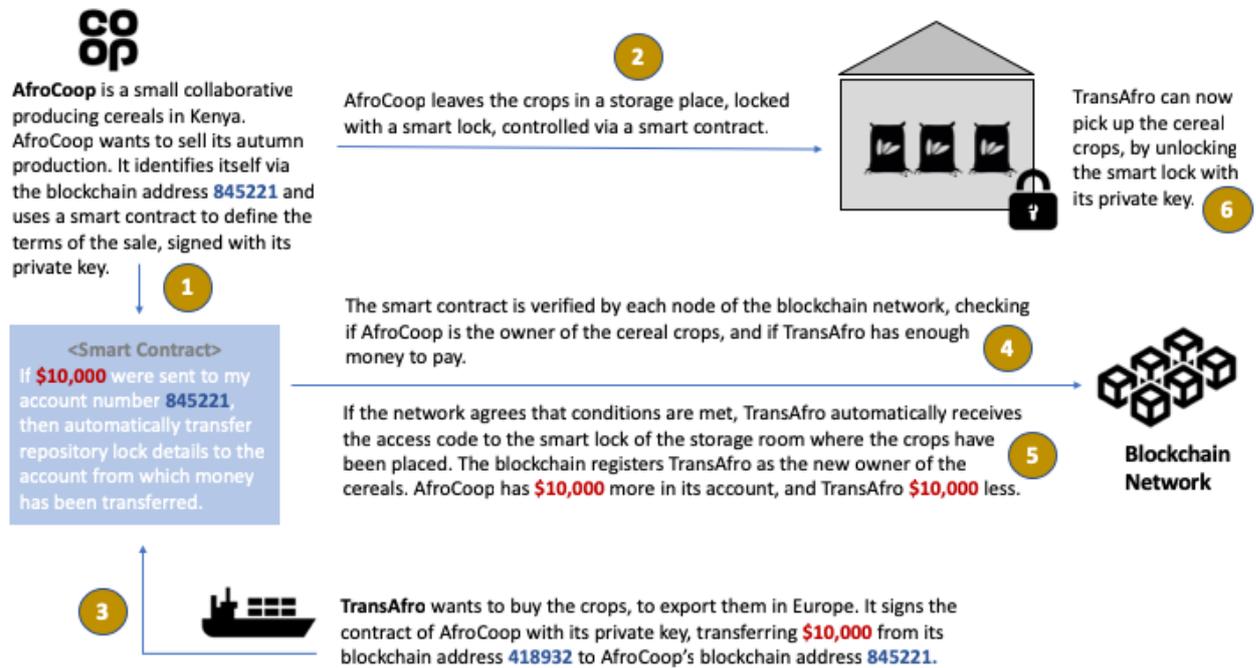

Figure 3: An example demonstrating how a smart contract could be executed in 6 steps, for automating and enhancing trust in transactions involving small farmers and cooperatives of small farmers.

Blockchain could also facilitate insurance programs for securing farmers (i.e. members of the cooperatives) against unpredicted weather conditions that affect their crops or other risks such as natural disasters (Jha, Andre and O. 2018). The idea behind the ARBOL project is via customized agreements, farmers can receive payments for droughts, floods, or other adverse weather outcomes that negatively affect their crop (ArbolMarket 2019).

## 3.5 Waste reduction and environmental awareness

Various waste management initiatives have incorporated blockchain technology. Worth mentioning is the Plastic Bank (Plastic Bank 2019), a global recycling venture founded in Canada to reduce plastic waste in developing countries – so far Haiti, Peru and Colombia, with plans to extend this year to Indonesia and Philippines. The initiative rewards people who bring plastic rubbish to bank recycling centres, and this reward is provided via blockchain-secured digital tokens. With these tokens, people can purchase things like food or phone-charging units in any store, using the Plastic Bank app (Steenmans and Taylor 2018). The Plastic Bank initiative seems to be successful till date, with more than one million participants, more than 2,000 collector units and three million kilograms of plastic collected in Haiti since 2014. A company with a mission similar to Plastic Bank is the Agora Tech Lab (Agora Tech Lab 2018), aiming to promote circular economy initiatives by rewarding responsible behavior.

Another example of the use of blockchain technology is emerging in railway stations. Waste management in French stations has traditionally been chaotic, with hundreds of tones of waste produced each year. A system developed by SNCF subsidiary Arep uses blockchain to allow detailed information to be collected, using Bluetooth to continually update on quantities of each type of waste, which waste managers collected it and how it is being moved around (SNCF 2017). Blockchain is used to record any actions taken and the overall collection process.

Other commercial solutions using blockchain to improve recycling and sorting of waste produced along the food chain include Recereum (Recereum 2017) and Swachhcoin (Swachhcoin 2018).

Finally, blockchain can help to raise awareness about the environmental characteristics of the food produced. A crucial problem here is the degradation of land, soil and water where food is being produced. In particular, the quality of soil is important towards the realization of the United Nations Sustainable Development Goals (SDG) (Keesstra, et al. 2016). In this context, the sustainable development, proper management and rational use of agricultural fields, water resources and soils is of utmost importance (Keesstra, et al. 2018). Tracing this information via the supply chain, making it visible to the public, is essential for putting public pressure to producers and policy-makers on the aspect of how the food is produced in a sustainable manner.

**3.6 Supervision and management**

Blockchain technology can also be harnessed as a credit evaluation system to strengthen the effectiveness of supervision and management in the food supply chain. It can also be used to improve the monitoring of international agreements relevant to agriculture, such as World Trade Organization agreements and the Paris Agreement on Climate Change (Tripoli and Schmidhuber 2018). The authors in (Mao, et al. 2018) have developed a system, based on the Hyperledger blockchain, which gathers credit evaluation text from traders by smart contracts on the blockchain. Traders' credit can then be used as a reference for regulators, to assess their credibility. By applying blockchain, traders can be held accountable for their actions in the process of transaction and credit evaluation by the regulators. As another example, AgriBlockIoT is a fully decentralized, blockchain-based solution for agri-food supply chain management (Caro, Ali, et al. 2018), able to seamless integrate IoT devices producing and consuming digital data along the chain. A similar research effort, combining IoT sensors and cloud technologies was proposed in (Davcev, et al. 2018), targeting the management of a grape farm near the City of Skopje, North Macedonia.

Blockchain-based contracts can also mitigate the exploitation of labour in agriculture, protecting workers with temporary agreements and employment relationships in the agricultural sector (Pinna and Ibba 2018). When labour agreements become part of the blockchain, it is easier for the authorities to control fairness in payments and also taxation. Coca-Cola has attempted to employ blockchain to sniff out forced labor in the sugarcane sector (Gertrude Chavez-Dreyfuss, Reuters 2018).

Quality measurement and monitoring are also relative aspects, where quality assurance is defined as the avoidance of failures such as delays to final destinations, poor monitoring, and frauds, as well as the assurance that the quality of the products (e.g. crops, meat, dairy) is maintained good along the transfer through the food chain, i.e. good storing conditions, no contamination or impurities etc. Several properties defining a good quality of grains are listed in (Brooker, Bakker-Arkema and Hall 1992). The preliminary results in (Lucena, et al. 2018) support a potential demand for a blockchain-based certification, which would lead to an added valuation of its selling price around 15% for genetically modified (GM)-free soy in the scope of a business network for grain exports in Brazil. This added valuation would be the outcome of more reliable and efficient quality assurance process on the grains, facilitated by blockchain. Blockchain was also used to record events taking place in the rice value chain, ensuring the security and quality of rice in the transportation process (Kumar and Iyengar 2017).

Finally, blockchain could be used to manage common resources such as energy and water, prevent speculation in the trading of these resources (Poberezhna 2018).

4. Analysis of the Findings

Table 1 shows blockchain technology initiatives/projects, in relation to the goods and/or products targeting, based on the examples presented in Sections 3.1-3.6. The last column indicates the objectives for employing blockchain technology at each case. Financial reasons are associated with food traceability in the commercial initiatives. As the table indicates, pilot studies have been implemented in a wide range of different products or at the food supply system as a whole. Some research-oriented studies examined the use of blockchain together with emerging technologies such as IoT, RFID, NFC, QR codes etc., focusing on automation of production and more productivity and transparency (Tian 2016), (Kim, et al. 2018), (Boehm, Kim and Hong 2017).

| Goods, Products, Resources | Initiative/Project/Company Involved | Objectives |
|---|---|---|
| Soybeans | LDC (Hoffman and Munsterman 2018) | Financial, Faster Operations |
| Grains | AgriDigital (AgriDigital 2017), GEBN study (Lucena, et al. 2018) | Financial, Supervision and management |
| Olive oil | OlivaCoin (OlivaCoin 2016) | Financial, Small farmers support |
| Turkeys | Cargill Inc. (Bunge 2017), Hendrix Genetics (Hendrix Genetics 2018) | Traceability, Animal welfare |
| Mangoes | Walmart, Kroger, IBM (CB Insights 2017), (Kamath 2018), Nestle (ITUNews 2018) | Traceability |
| Canned pumpkin | Nestle (ITUNews 2018) | Traceability |
| Pork | Walmart, Kroger, IBM (CB Insights 2017), (Kamath 2018) | Traceability |
| Sugar cane | Coca-Cola (Gertrude Chavez-Dreyfuss, Reuters 2018) | Supervision and Management |
| Beer | Downstream (Ireland Craft Beers 2017) | Traceability |
| Beef | "Paddock to plate" project (Campbell 2017), BeefLedger (BeefLedger Limited 2017), JD.com (Adele Peter, Fast Company 2017) | Traceability |
| Cannabis | Medical Cannabis Tracking (MCT) system (Abelseth 2018) | Traceability |
| Chicken | Gogochicken (Adele Peter, Fast Company 2017), Grass Roots Farmers Cooperative (Grass Roots Farmers' Cooperative 2017), OriginTrail (OriginTrail 2018) | Traceability |
| Wood (Chestnut trees) | Infotracing (Figorilli, et al. 2018) | Traceability |
| Sea-food | Intel (Hyperledger 2018), WWF (WWF 2018), Balfegó (Balfegó Group, 2017) | Environmental impact, Traceability |
| Table grapes | "Blockchain for agrifood" project (Ge, et al. 2017), Grape farm near the City of Skopje (Davcev, et al. 2018) | Experimental feasibility study, Supervision and management |
| Organic food | Soil Association Certification (Soil Association Certification 2018) | Financial, Traceability, Small farmers support |
| Food waste | Plastic Bank (Plastic Bank 2019), Agora Tech Lab (Agora Tech Lab 2018), SNCF (SNCF | Waste rediction |

|  | 2017), Recereum (Recereum 2017), Swachhcoin (Swachhcoin 2018) |  |
|---|---|---|
| Water | Global water assets (Poberezhna 2018) | Supervision and management |
| Rice | Quality of rice in transportation (Kumar and Iyengar 2017) | Supervision and management |
| Food chain in general | AgriLedger (AgriLedger 2017), FarmShare (FarmShare 2017), Carrefour (Carrefour 2018), ripe.io (Ripe.io 2017), OriginTrail (OriginTrail 2018), (AgriBlockIoT (Caro, Ali, et al. 2018), Food supply chain prototypes enhanced with other technologies (Tian 2017), (Kim, et al. 2018), (Boehm, Kim and Hong 2017). | Financial, Traceability, Food safety, Small farmers support, Waste reduction, Supervision and management |

Table 1: Goods and products, in relation to projects using blockchain technology and their overall objectives.

It is interesting to see the underlying technology used by the 49 different projects, initiatives and papers identified through this survey, to empower blockchain-based transactions. The most popular technology adopted was Ethereum (9 projects/initiatives, 18%), followed by Hyperledger Fabric (6 projects/initiatives, 12%). Seven projects preferred to develop their own blockchain solution (Zeto 2018), (Carrefour 2018), (JD.com Blog 2018), (Ripe.io 2017), (OriginTrail 2018), (OlivaCoin 2016), (AgriDigital 2017). From the other initiatives, BigchainDB was employed in (Tian 2017), the Bitcoin protocol in (Bunge 2017), BeefLedger in (Campbell 2017), the ZhongAn blockchain open platform in (Adele Peter, Fast Company 2017), Provenance in (Grass Roots Farmers' Cooperative 2017), (Soil Association Certification 2018), Hyperledger Sawtooth in (Hyperledger 2018), the Azure Blockchain Workbench together with Ethereum in (Figorilli, et al. 2018) and, finally, a combination of Ethereum and Hyperledger Sawtooth (Caro, Ali, et al. 2018). The remaining 17 projects (34%) did not reveal any information about the underlying structure of their blockchain-based solutions.

Figure 4 depicts the maturity level of the related work as identified through this survey, starting from conceptual stage (10 projects/initiatives, 20%) up to full integration to normal operations of the entity involved (4 projects/initiatives, 8%). As the figure shows, the majority of the projects are either in implementation phase (13 projects/initiatives, 26.5%) or in a proof-of-concept stage, through small pilot studies (14 projects/initiatives, 28.5%). Research-based projects tend to reach the level of a small pilot study only, most of them being at a conceptual or implementation stage. All 8 projects/initiatives (16%) that develop large-scale case studies are supported and ran by big companies. With large-scale studies we refer to hundreds of thousands of goods/products involved, interaction with thousands of consumers, and/or involvement of tens to hundreds of intermediate actors in the supply chain. The fact that only 4 initiatives have reached the phase of a complete integration to normal operations (AID Tech 2017), (Blockchain for Zero Hunger 2017), (Ireland Craft Beers 2017), (Plastic Bank 2019), indicates that blockchain technology is still being studied by companies and organizations, perceived mostly as an experimental new tool and as an emerging technology with certain potential. It is also likely that companies perform pilot studies involving blockchain for marketing reasons (due to the hype of this technology) or for the possibility of a competitive advantage in the future.

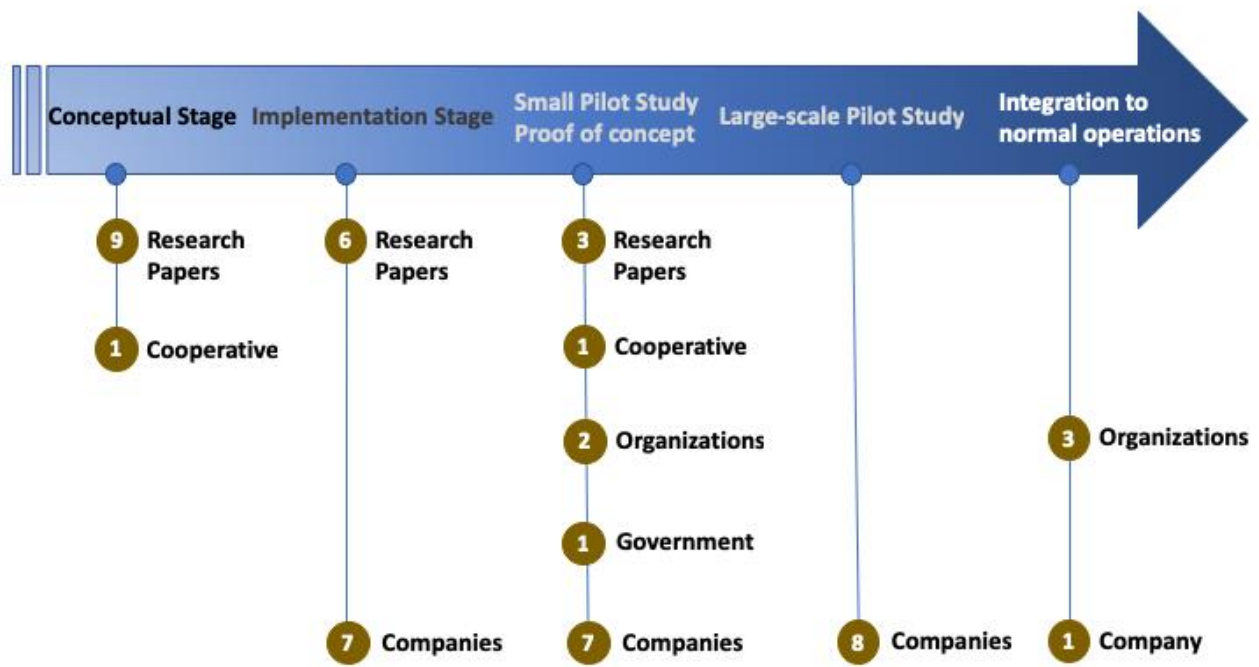

Figure 4: Maturity level and number of projects, initiatives and research papers as identified in this study.

Finally, it is worth investigating whether the aforementioned projects/initiatives are still running, or whether they have stopped and/or failed. This would be a key indicator of the economic viability of blockchain-empowered projects. Unfortunately, it is hard to address this question because most of the initiatives started quite recently, i.e. in 2016 (7 projects/initiatives, 14%), in 2017 (12 projects/initiatives, 24.5%), or in 2018 (22 projects/initiatives, 45%). Due to the small lifetime, most projects are on-going and this makes their assessment difficult.

Based on our research, taking into account updates about each project, news articles and any other recorded activity, we suspect that 7 out of the 29 commercial initiatives (24%, including governmental and NGO-based ones, excluding research papers) might have become inactive. These initiatives are the following: (AID Tech 2017), (Jha, Andre and O. 2018), (Grass Roots Farmers' Cooperative 2017), (WWF 2018), (Balfegó Group 2017), (FarmShare 2017), (Soil Association Certification 2018), (SNCF 2017), (Recereum 2017). This percentage of possible fallouts is definitely large, and it might be an indication of the overall complexity of the blockchain technology, or the immaturity of the market for complete integration to companies' everyday operations. It could also show that some companies/organizations have finished their pilots, and they are still studying the possibility of massive adoption. Time will show if the latter is the case.

## 5. Potential Benefits

Blockchain technology offers many benefits, as it can provide a secure, distributed way to perform transactions among different untrusted parties (Yuan, et al. 2019), (Pearson, et al. 2019), (Creydt en Fischer 2019). This is a key element in agriculture and food supply chains, where numerous actors are involved from the raw production to the supermarket shelf (Lin, et al. 2017), (Tripoli and Schmidhuber 2018). To improve traceability in value chains, a decentralized ledger helps to connect inputs, suppliers, producers, buyers, regulators that are far apart, who are under different programs, different rules (policies) and/or using different applications (Lee, et al. 2017). Via smart contracts, manufacturers can develop scalable and flexible businesses at a lower cost, and the overall effectiveness of manufacturing services can be improved (Li, et al. 2018).

Blockchain has the potential to monitor social and environmental responsibility, improve provenance information, facilitate mobile payments, credits and financing, decrease transaction fees, and facilitate real-time management of supply chain transactions in a secure and trustworthy way (Lee, et al. 2017). In the case of an outbreak of an animal or plant disease, contaminated products could be traced more quickly (Tripoli and Schmidhuber 2018). Blockchain could even be used to make agricultural robotic swarm operations more secure, autonomous and flexible (Ferrer 2018).

In particular, blockchain seems very suitable to be used in the developing world, as we saw in the Section 3.4, in relation to small farmers' support. Other scenarios could involve finance and insurance of rural farmers (Chinaka 2016), as well as facilitation of transactions in developing countries. Cash transactions lack traceability, which ultimately hinders the ability of small- and medium-sized enterprises in developing countries, to access credit and new markets and to grow. The blockchain introduces a new method of accounting for value transfers that minimizes uncertainty and disintermediates the exchange of value with a decentralized and shared ledger, functioning as a digital institution of trust, with reduced (if any) transaction costs (Tripoli and Schmidhuber 2018). Although small farmers produce more than 80% of goods in developing countries, in most cases they do not have support of services such as finance and insurance (Chinaka 2016). Blockchain could also be used to fight corruption and the insufficient environmental, social and economic regulatory frameworks in these countries (Rejeb 2018). More examples on how blockchain could help empowering the poor in developing countries are listed in (Thomason, et al. 2018), with focus on tracking climate finance, results tracking, climate adaptation, financial inclusion, and identity.

Concerning the developed world, existing problems such as unfair pricing and the influence of big companies have historically limited the environmental/economic sustainability of smaller farms. Blockchain could help in a fairer pricing through the whole value chain. An example of how blockchain could be used for record keeping of water quality data along a catchment area is discussed in (IWA 2018).

Moreover, the potential transparency provided by blockchains could facilitate the development of trading systems that are based on reputation. Reputation, as we have witnessed from various other trading systems where it has been used (e.g. eBay, Alibaba), improves the behavior of participating parties and increases their reliability, responsibility and commitment (Khaqqi, et al. 2018), (Sharma 2017).

Further, there is the potential benefit of increasing consumer awareness and empowerment, considering that the consumer is the market driving force. Consumer increased awareness would put pressure for more transparent, sustainable, safe and fair practices in food production. Since consumers are overwhelmed by the amount and complexity of certification labels, blockchain technology seems to have positive influences on consumers' purchasing decisions (Sander, Semeijn and Mahr 2018). Finally, the case study performed in (Perboli, Musso and Rosano 2018) shows that the cost of implementing a blockchain is highly sustainable when compared with the resulting benefits.

6. **Challenges and Open Issues**

There are various challenges for the wider adoption of blockchain technology, which are mentioned in related work under study and also in relevant survey and position papers *(Chang, Iakovou and Shi 2019)*, *(Galvez, Mejuto and Simal-Gandara 2018)*, *(Hald and Kinra 2019)*, *(Tribis, El Bouchti and Bouayad 2018)*, *(Zhao, et al. 2019)*, *(Pearson, et al. 2019)*. Table 2 lists potential benefits and existing barriers for the use of blockchain in agriculture and the food supply chain, as identified in Section 5 and Sections 6.1-6.5 respectively, as well as in *(Chang, Iakovou and Shi 2019)*, *(Pearson, et al. 2019)*. A case study in the Netherlands revealed that SME lack the required size, scale or know-how needed, in order to invest in blockchain by themselves *(Ge, et al. 2017)*.

| **Opportunities and potential benefits** | **Challenges and barriers** |
|---|---|
| Traceability in value chains | SME have difficulties in adopting the technology |
| Support for small farmers | Information infrastructure might prevent access to markets for new users |
| Finance and insurance of rural farmers | Lack of expertise by small SME |
| Facilitation of financial transactions in developing countries | High uncertainties and market volatility |
| Fairer pricing through the whole value chain | Limited education and training platforms |
| A useful platform in emission reduction efforts | No regulations in place |
| Consumer awareness and empowerment | Lack of understanding among policy makers and technical experts |
| More informed consumer purchasing decisions | Open technical questions and scalability issues (e.g. latency of transactions) |
| Increased sustainability and reduction of waste | Digital divide among developed and developing world |
| Reduced transaction fees and less dependence on intermediaries | Decline of cryptocurrencies in market share and high volatility (reputation issues) |
| More transparent transactions and less frauds | Cost of computing/IoT equipment required |
| Better quality of products, lower probability for foodborne diseases | Design decisions might reduce overall flexibility |
| | Privacy issues |
| | Some quality parameters of food products cannot be monitored by objective analytical methods, especially environmental indicators |

Table 2: Potential benefits and existing barriers for the use of blockchain in agriculture.

### 6.1 Accessibility

Blockchain needs to become more accessible and this is a big challenge considering that the underlying digital technology can become increasingly complex, as more components are integrated into blockchain (IoT, RFID, sensors and actuators, robots, biometric data, big data etc.) (Tian 2016), (Figorilli, et al. 2018), (Kim, et al. 2018), (Rabah 2018). In fact, in order to be functional, blockchains must rely on external systems to obtain accurate information from the real-world. These are the so-called *oracles* that connect the physical and digital worlds, and usually come from automated sensor readings (i.e. hardware oracles), datasets from the web applications (i.e. software oracles), and manual records (i.e. human oracles). However, the necessity of such third-party intermediaries might compromise the blockchain building of decentralized trust. Substantial research is being carried out on how to tackle the *oracle problem* in blockchain, particularly for finance and smart contract-related applications. The proposed solutions generally rely on developing decentralized and consensus-based oracle solutions, and novel methods of authenticating oracle data.

While blockchains can connect complex global supply chains, the information infrastructure required to operate and maintain the system might prevent access to markets for new users or food suppliers. The systems could, in effect, become a technical barrier to trade, thus reducing market competition and access (Pearson, et al. 2019).

Moreover, there is a general lack of awareness and skills on blockchain technology (Zhao, et al. 2019), while training platforms are still limited (ICT4Ag 2017). Besides policy-makers, capacitation on the blockchain technology is also fundamental for the food value chain stakeholders. Conceptual metaphors for understanding and accepting blockchain are discussed in (Swan and De Filippi 2017). Various startups have been working in developing software to make blockchain technology easier for farmers to use, such as 1000 EcoFarms (1000EcoFarms 2017), which has aggregated all the important blockchain processes relevant to food, farming and agriculture, using FoodCoin as the proposed ecosystem (FoodCoin 2017). OriginChain is a software system that restructures the current central database systems with blockchain (Xu, et al. 2019).

**6.2 Governance and Sustainability**

Despite the rather long list of initiatives presented in this review, convincing business cases are still scarce, due to large number of uncertainties involved and the early stages of the technology. This observation was made also in a relevant survey (Tribis, El Bouchti and Bouayad 2018). Hence, the long-term impact of blockchain on governance, economic sustainability, and on social aspects still needs to be assessed. Some authors have pointed out that an excess of information transparency and the immutability of the data stored in blockchains might bring new challenges for the performance of supply chains (Hald and Kinra 2019). On the one hand, permanent data visibility might compromise privacy issues and could eventually strengthen the surveillance power of centralized entities. On the other hand, large corporations might implement private and permissioned blockchains that could underpin oligopolistic practices (Pearson, et al. 2019).

Paradoxically, blockchain has also been described as a potentially *deskilling* technology for workers and organizations (Hald and Kinra 2019). The increased automation of tasks and procedures throughout supply chains and the elimination of transaction intermediaries might reduce significantly the human intervention, with the consequent loss of skilled jobs. The margin for human intervention in blockchain-managed supply chains could be reduced significantly. However, we must consider that such phenomena have occurred in all previous technological revolutions, which have in turn demanded new skills and capacities at the labor market.

Finally, it is worth adding that the quality parameters of food products (being more transparent to the consumer by means of the blockchain) justify in many cases higher prices. Therefore, they are often in the focus of food fraudsters (Creydt en Fischer 2019), thus governance is important also in this aspect.

**6.3 Regulation**

Policy development and regulation in relation to blockchain practices is both a necessity and an important barrier for its wider adoption (Zhao, et al. 2019), (Pearson, et al. 2019). As cryptocurrencies form the most complete to date global blockchain study case (Yli-Huumo, et al. 2016), the current experience of analyzing these cryptocurrencies indicates that they are vulnerable to speculators and their price has large fluctuations almost daily. The recent decline in market share and high volatility of the financial value of the most popular cryptocurrencies reduces the overall trust of the public in the underlying blockchain technology of cryptocurrencies, thus having a negative psychological effect on its reputation (Gaurav 2019). Hence, without some form of regulation, cryptocurrencies are not trustful to be used yet in food supply chains as a complete solution. The absence of regulation makes this problem persistent.

A lack of (common) understanding among policy makers and technical experts still exists on how blockchain technology and transactions based on some currency should be used (ICT4Ag 2017).

## 6.4 Technical Challenges and Design Decisions

There are many design decisions that affect the existing blockchains or the ones under development (e.g. (AgriDigital 2017), (AgriLedger 2017), (FarmShare 2017), (Ripe.io 2017), (OriginTrail 2018)). For example, shall they be permissioned (i.e. participants are trusted), permission-less, open (i.e. everyone can join) or closed systems (Jayachandran 2017)? Who should own the blockchain (Pearson, et al. 2019)? Observing the existing permission-less blockchains, the latency of transactions might be several minutes up to some hours to finish, until all participants update their ledgers and the smart contracts become publicly accessible. Such design decisions affect the operation of the blockchain system and this creates some lack of flexibility which, under certain circumstances, might make blockchain solutions less efficient than the equivalent conventional centralized approaches.

Moreover, some of the quality parameters of food products can be monitored by objective analytical methods, but not all of them (Creydt en Fischer 2019). Some parameters, especially environmental ones (Keesstra, et al. 2018), are difficult to include, assess and audit.

Further, existing blockchain protocols face serious scalability obstacles (Eyal, et al. 2016), (Pearson, et al. 2019), since the current processing of transactions is limited by parameters such as the size and interval of the transaction block (Tribis, El Bouchti and Bouayad 2018). The majority of the proposed blockchain-based frameworks were only tested on a limited scale in a "laboratory" environment. Although blockchain offers advanced security, there are high risks related to loss of funds, just because the account owner might have lost accidentally the private keys needed to access and manage the account.

Finally, privacy issues are important (Zhao, et al. 2019). Since every transaction is recorded on a common ledger, users can be identified by their public keys. Although this aspect ensures transparency and helps to build trust, at the same time it does not protect users' privacy. This privacy is particularly important in the food supply ecosystem, since many actors are competitors with each other. Thus, maintaining a certain level of privacy is an existing challenge of blockchain technologies. Various methods for privacy protection in blockchain systems are discussed in the survey of (Feng, et al. 2018).

## 6.5. Digital Gap Between Developed and Developing Countries

As mentioned in the previous subsection, the farmers need to effectively understand blockchain before adopting it (Tribis, El Bouchti and Bouayad 2018). However, the priority for farmers in many parts of the world is subsistence, so that they need to dedicate their efforts in farming and have no expertise in cutting edge technologies. Since blockchain technologies require a high degree of computing equipment (i.e. in some blockchain systems, such as permissionless ones) (Zhao, et al. 2019), it is difficult to find these resources in developing countries. Hence, there seems to be a gap among the developed and developing world, in respect to digital competence and access to the blockchain technology (Maru, et al. 2018). Many of the bibliographic sources come from developed countries with a well-organized and wealthy primary sector (i.e. the USA, Australia, Europe, etc.). This digital divide was also observed in the use of big data in agriculture (Kamilaris, Kartakoullis and Prenafeta-Boldú 2017). Some authors do make the important observation that most of the current projects are in developed countries, but no significant questions are raised around this in their conclusion. Since blockchain is being constantly referred to as solving many developing world challenges, asking '*why the gap?*' is an important question, and a legitimate area for future research. Figure 5 illustrates the number of blockchain experiences at the public sector in various countries around the world (Killmeyer, White and Chew 2017).

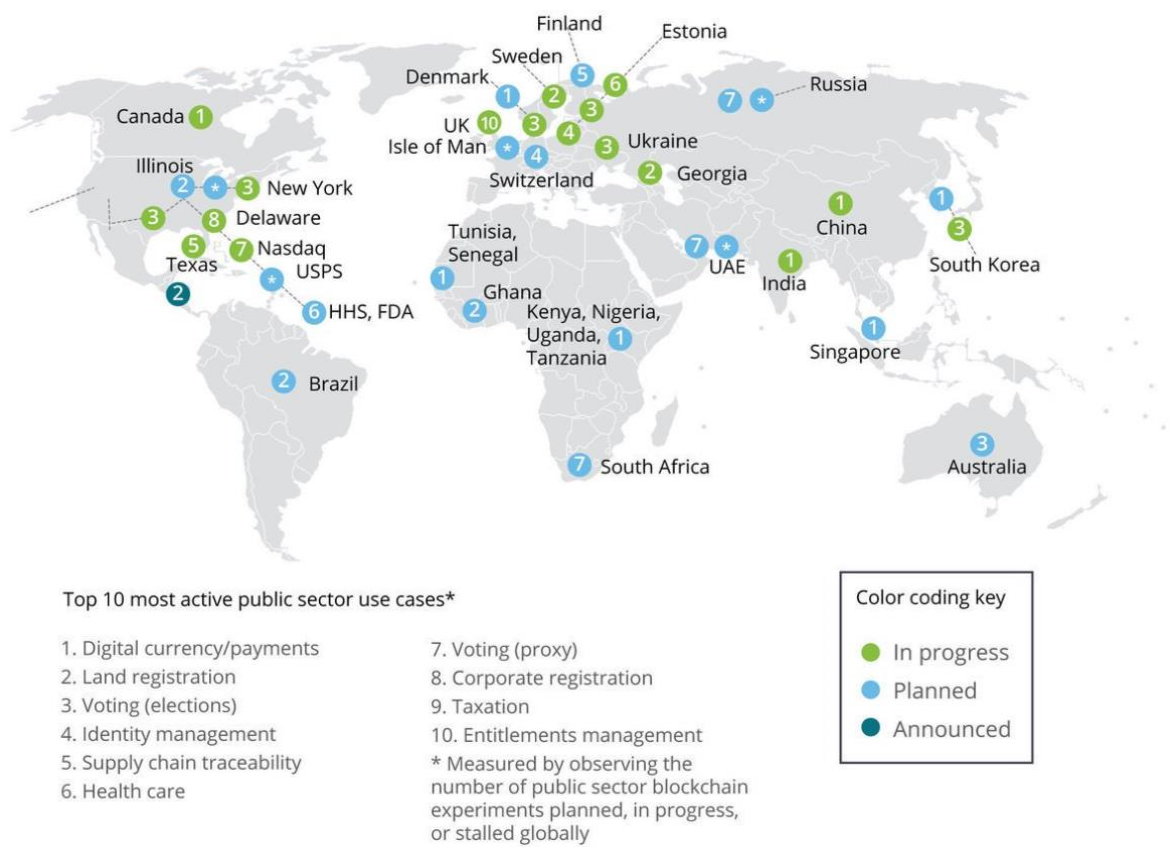

Figure 5: Blockchain in the public sector in 2017 (Source: *(Killmeyer, White and Chew 2017)*, appropriate permissions have been obtained from the copyright holders of this work).

It seems indeed that most of the on-going experiments happen in developed regions. Considering that blockchain might be an important opportunity for small farmers (see Section 3.4 and Section 5), developmental aid should focus on training and technology transfer to the farmers in developing areas with the view of bringing actual solutions to the specific conditions that restrain their socioeconomic progression.

## 7. Conclusion

This article demonstrates that blockchain technology is already being used by many projects and initiatives, aiming to establish a proven and trusted environment to build a transparent and more sustainable food production and distribution, integrating key stakeholders into the supply chain. Yet, there are still many issues and challenges that need to be solved, beyond those at technical level.

To reduce barriers of use, governments must *lead by example* and foster the digitalization of the public administration. They should also invest more in research and innovation, as well as in education and training, in order to produce and demonstrate evidence for the potential benefits of this technology. Gupta (Gupta 2017) discusses the possible transition of governments towards the use of the blockchain, noting the fact that governments and their relevant departments should observe and understand the particular "pain points", addressing them accordingly.

From a policy perspective, various actions can be taken, such as encouraging the growth of blockchain-minded ecosystems in agri-food chains, supporting the technology as part of the general goals of optimizing the competitiveness and ensuring the sustainability of the agri-food supply chain, as well as designing a clear regulatory framework for blockchain implementations.

The economic sustainability of the existing initiatives, as they have been presented in this paper, still needs to be assessed and the outcomes of these economic studies are expected to influence the popularity of the blockchain technology in the near future, applied in the food supply chain domain.

Summing up, blockchain is a promising technology towards a transparent supply chain of food, but many barriers and challenges still exist, which hinder its wider popularity among farmers and food supply systems. The near future will show if and how these challenges could be addressed by governmental and private efforts, in order to establish blockchain technology as a secure, reliable and transparent way to ensure food safety and integrity. It is very interest to see how blockchain will be combined with other emerging technologies (big data, robotics, IoT, RFID, NFC, hyperspectral imaging etc.), towards higher automation of the food supply processes, enhanced with full transparency and traceability.

**Acknowledgments**

This research has been supported by the P-SPHERE project, which has received funding from the European Union's Horizon 2020 research and innovation programme under the Marie Skodowska-Curie grant agreement No 665919. The support of the CERCA Program and of the Consolidated Research Group TERRA (ref. 2017 SGR 1290), both from the Generalitat de Catalunya, is also acknowledged.